\documentclass[12pt,preprint]{aastex}
\usepackage{graphicx}
\usepackage{lscape}
\usepackage{enumerate}
\usepackage{xcolor}

\newcommand{\pivec}{\mbox{\boldmath $\pi$}}

\lefthead{JUNG ET AL.}
\righthead{ }

\begin{document}
\title{
OGLE-2018-BLG-0567Lb and OGLE-2018-BLG-0962Lb: Two Microlensing Planets through Planetary-Caustic Channel
}

\author{
Youn~Kil~Jung$^{1,2,28}$, 
Cheongho~Han$^{3,28}$,
Andrzej~Udalski$^{4,29}$,
Andrew~Gould$^{1,5,6,28}$,
Jennifer~C.~Yee$^{7,28}$, \\
and \\
Michael~D.~Albrow$^{8}$, Sun-Ju~Chung$^{1,2}$, Kyu-Ha~Hwang$^{1}$, 
Yoon-Hyun~Ryu$^{1}$, In-Gu~Shin$^{1}$, Yossi~Shvartzvald$^{9}$, Wei~Zhu$^{10}$,
Weicheng~Zang$^{11}$, Sang-Mok~Cha$^{1,12}$, Dong-Jin~Kim$^{1}$, Hyoun-Woo~Kim$^{1}$, 
Seung-Lee~Kim$^{1,2}$, Chung-Uk~Lee$^{1,2}$, Dong-Joo~Lee$^{1}$, Yongseok~Lee$^{1,12}$, 
Byeong-Gon~Park$^{1,2}$, Richard~W.~Pogge$^{5}$ \\
(The KMTNet Collaboration) \\
Przemek~Mr{\'o}z$^{4,13}$, Micha{\l}~K.~Szyma{\'n}ski$^{4}$, Jan~Skowron$^{4}$, Radek~Poleski$^{4,5}$,
Igor~Soszy{\'n}ski$^{4}$, Pawe{\l}~Pietrukowicz$^{4}$, Szymon~Koz{\l}owski$^{4}$, Krzystof~Ulaczyk$^{14}$, 
Krzysztof~A.~Rybicki$^{4}$, Patryk~Iwanek$^{4}$, Marcin~Wrona$^{4}$ \\
(The OGLE Collaboration) \\
}
\bigskip\bigskip
\affil{$^{1}$Korea Astronomy and Space Science Institute, Daejon 34055, Republic of Korea}
\affil{$^{2}$University of Science and Technology, Korea, 217 Gajeong-ro Yuseong-gu, Daejeon 34113, Korea}
\affil{$^{3}$Department of Physics, Chungbuk National University, Cheongju 28644, Republic of Korea}
\affil{$^{4}$Warsaw University Observatory, Al. Ujazdowskie 4, 00-478 Warszawa, Poland}
\affil{$^{5}$Department of Astronomy, Ohio State University, 140 W. 18th Ave., Columbus, OH 43210, USA}
\affil{$^{6}$Max-Planck-Institute for Astronomy, K$\rm \ddot{o}$nigstuhl 17, 69117 Heidelberg, Germany}
\affil{$^{7}$Center for Astrophysics $|$ Harvard \& Smithsonian, 60 Garden St., Cambridge, MA 02138, USA}
\affil{$^{8}$University of Canterbury, Department of Physics and Astronomy, Private Bag 4800, Christchurch 8020, New Zealand}
\affil{$^{9}$Department of Particle Physics and Astrophysics, Weizmann Institute of Science, Rehovot 76100, Israel}
\affil{$^{10}$Canadian Institute for Theoretical Astrophysics, University of Toronto, 60 St George Street, Toronto, ON M5S 3H8, Canada}
\affil{$^{11}$Department of Astronomy, Tsinghua University, Beijing 100084, China}
\affil{$^{12}$School of Space Research, Kyung Hee University, Yongin 17104, Republic of Korea}
\affil{$^{13}$Division of Physics, Mathematics, and Astronomy, California Institute of Technology, Pasadena, CA 91125, USA}
\affil{$^{14}$Department of Physics, University of Warwick, Gibbet Hill Road, Coventry, CV4 7AL, UK}
\footnotetext[28]{The KMTNet Collaboration.}
\footnotetext[29]{The OGLE Collaboration.}

\begin{abstract}
We present the analyses of two microlensing events, OGLE-2018-BLG-0567 and OGLE-2018-BLG-0962. 
In both events, the short-lasting anomalies were densely and continuously covered by two high-cadence 
surveys. The light-curve modeling indicates that the anomalies are generated by source crossings 
over the planetary caustics induced by planetary companions to the hosts. The estimated planet/host 
separation (scaled to the angular Einstein radius $\theta_{\rm E}$) and mass ratio are 
$(s, q) = (1.81, 1.24\times10^{-3})$ and $(s, q) = (1.25, 2.38\times10^{-3})$, respectively. 
From Bayesian analyses, we estimate the host and planet masses as 
$(M_{\rm h}, M_{\rm p}) = (0.24_{-0.13}^{+0.16}\,M_{\odot}, 0.32_{-0.16}^{+0.34}\,M_{\rm J})$ and 
$(M_{\rm h}, M_{\rm p}) = (0.55_{-0.29}^{+0.32}\,M_{\odot}, 1.37_{-0.72}^{+0.80}\,M_{\rm J})$, respectively. 
These planetary systems are located at a distance of $7.07_{-1.15}^{+0.93}\,{\rm kpc}$ 
for OGLE-2018-BLG-0567 and $6.47_{-1.73}^{+1.04}\,{\rm kpc}$ for OGLE-2018-BLG-0962, 
suggesting that they are likely to be near the Galactic bulge. The two events prove 
the capability of current high-cadence surveys for finding planets through the 
planetary-caustic channel. We find that most published planetary-caustic planets 
are found in Hollywood events in which the source size strongly contributes to 
the anomaly cross section relative to the size of the caustic.
\end{abstract}

\section{Introduction}

The signature of a microlening planet is almost always a short-lasting anomaly in the smooth 
and symmetric lensing light curve produced by the host of the planet. In principle, the signature 
can appear at any position of the lensing light curve \citep{gaudi12}. In reality, however, 
the signatures of planets detected in the earlier phase of lensing experiments appeared mainly 
near the peak of lensing light curves.

The bias toward central anomalies is mostly attributed to the limitation of early lensing surveys. 
With a roughly 1 day cadence of the first-generation survey experiments, e.g., the MACHO \citep{alcock95}, 
OGLE-I \citep{udalski92}, MOA-I \citep{bond01} surveys, it was difficult to detect planetary signals 
lasting of order $1$ day or less by the survey experiments. To meet the cadence requirement for planet 
detections, \citet{gould92b} proposed an observational mode, in which wide-field surveys with a low 
cadence monitor a large area of sky mainly to detect lensing events, and followup experiments conduct 
high-cadence observations for a small number of lensing events detected by the surveys using a network 
of multiple narrow-field telescopes. However, this mode of observations had the drawback that only 
a handful of lensing events could be monitored by followup observations. Combined with the low probability 
of planetary perturbations, this implied a low planet detection rate for this phase of the experiments. 
In fact, for the first several years, there were no securely detected planets using this mode, 
although there was one tentative detection in the event MACHO 98-BLG-35 \citep{rhie00}.

The first three microlensing detections were found using the survey+followup strategy. 
In the first event, OGLE 2003-BLG-235/MOA 2003-BLG-53 \citep{bond04}, the planet was 
found by the surveys, but the MOA survey carried out additional followup observations 
in response to the planetary anomaly. The next two planets, OGLE-2005-BLG-071Lb \citep{udalski05} 
and OGLE-2005-BLG-390Lb \citep{beaulieu06}, were both found through extensive followup 
observations of known microlensing events that were initiated before the planetary 
anomaly began. The discovery of OGLE-2005-BLG-071Lb provided a practical lesson in 
the value of high-magnification events \citep{griest98} for detecting planets 
through followup observations.

In the following years, the planet detection rate using the survey+followup mode was 
substantially increased by focusing on events with very high magnifications. Several 
factors contributed to the increase of the detection rate. First, the planet detection 
efficiency for high-magnification events is high. This is because a planet located in 
the lensing zone of its host always induces a small central caustic near the position of 
the host, and the trajectory of a high-magnification event passes close to the central caustic. 
This yields a high probability that a planet will produce a perturbation and also confine 
that perturbation to a short duration of time while the event is highly magnified, not throughout 
the whole event. As a result, the time of the planetary signal, i.e., the peak of the light curve, 
can be predicted in advance and enable one to efficiently use resources for followup observations. 
By contrast, predicting the time of a planetary signal through other channels is difficult. 
Finally, highly magnified source stars are bright enough to be observed with small-aperture telescopes, 
down to sub-meter amateur-class telescopes, and this enables one to maximize available telescopes 
for followup observations, e.g., OGLE-2005-BLG-071 \citep{udalski05,dong09}. Thus, the planets 
detected from the survey+followup experiments were detected mainly through the high-magnification channel, 
and this led to the bias toward central caustic perturbations.

The current planetary lensing experiments are in the second phase, in which lensing events are 
observed by high-cadence surveys. The observational cadence of the lensing surveys in this phase 
has greatly increased with the employment of large-format cameras yielding very wide fields of view. 
The MOA experiment entered a new phase (MOA-II) by upgrading its instrument with a new wide-field 
camera composed of ten $2{\rm k} \times 4{\rm k}$ chips yielding a $2.2\,{\rm deg}^{2}$ field of 
view \citep{sumi13}. The OGLE survey is in its fourth phase (OGLE-IV) using $1.4\,{\rm deg}^{2}$ 
camera composed of 32 $2{\rm k} \times 4{\rm k}$ chips \citep{udalski15}. The KMTNet survey, 
which commenced its full operation in 2016 \citep{kim16}, utilizes three globally distributed 
telescopes, each of which has a camera with a $4.0\,{\rm deg}^{2}$ field of view. Being able to 
cover a large area of sky from a single exposure, the observational cadence of the current survey 
experiments now reaches $\Gamma\sim 4\,{\rm hr}^{-1}$ toward the dense bulge fields. 
This enables planet detections without additional followup observations.

With the operation of the high-cadence surveys, the detection rate of planets is rapidly increasing. 
One important reason for the rapid increase of the detection rate is that planets can be detected 
not only through the central-caustic channel but also through the additional planetary-caustic channel. 
Planets are detected through the planetary-caustic channel as anomalies produced by the 
source's approach close to the ``planetary caustic'', which denotes one of the two sets of planet-induced 
caustics lying away from the host. The planetary caustic lies at a position with a separation from the 
host of $s-1/s$, and thus planetary signals produced by this caustic can appear at any part of the 
lensing light curve depending on the planet-host separation $s$ (normalized to the angular Einstein 
radius $\theta_{\rm E}$). The planetary caustic is substantially larger than the central caustic, 
and thus the probability of a planetary perturbation is higher. Another importance of detecting planets 
through the planetary-caustic channel is that interpreting the planetary signal is usually not 
subject to the close-wide degeneracy \citep{griest98,dominik99}, which causes ambiguity in estimating 
the planet-host separations for most planets detected through the central-caustic channel.

In this paper, we present the analysis of two planetary microlensing events OGLE-2018-BLG-0567 
and OGLE-2018-BLG-0962, for which planets are both detected through a planetary-caustic channel. 
For both events, the signatures of the planets were densely and continuously covered by two 
high-cadence lensing surveys, and this leads us to unambiguously interpret the planetary signals.

\section{Observation}

The two planetary events were observed by the two lensing surveys conducted by the OGLE and KMTNet groups. 
The OGLE survey uses the $1.3\,{\rm m}$ telescope that is located at the Las Campanas Observatory in Chile. 
The KMTNet survey utilizes three $1.6\,{\rm m}$ telescopes that are located at the Siding Spring Observatory 
in Australia (KMTA), the Cerro Tololo Interamerican Observatory in Chile (KMTC), and the South African 
Astronomical Observatory in South Africa (KMTS). The global distribution of the KMTNet telescopes makes it 
possible to continuously monitor the events. In both surveys, observations were mainly conducted in the $I$-band, 
and a fraction of the images were taken in the $V$-band to determine the color of the microlensed source stars.

OGLE-2018-BLG-0567, $({\rm RA}, {\rm Dec})_{\rm J2000} = $(17:56:04.42, $-27$:59:13.6), or 
$(l, b) = (1^\circ\hskip-2pt .99,-1^\circ\hskip-2pt .49)$ in Galactic coordinates, was discovered 
on 2018 April 14 by the OGLE Early Warning System \citep[EWS:][]{udalski03}. The event was 
independently found by the KMTNet survey as KMT-2018-BLG-0890 from its event-finding algorithm \citep{kim18}. 
The observational cadence for the event is $\Gamma = 1\,{\rm hr}^{-1}$ for OGLE and 
$\Gamma = 2\,{\rm hr}^{-1}$ for KMTNet.

OGLE-2018-BLG-0962 was discovered by the OGLE EWS on 2018 June 2. It is located at 
$({\rm RA}, {\rm Dec})_{\rm J2000} = $(17:52:41.95, $-32$:18:33.3) or Galactic coordinates 
of $(l, b) = (-2^\circ\hskip-2pt .11,-3^\circ\hskip-2pt .04)$. The OGLE cadence for this 
direction is 3-10 times per night. The KMTNet collaboration also observed the event, 
with designation KMT-2018-BLG-2071, located in their two overlapping fields (BLG41 and BLG22). 
In combination, the KMTNet observations have a frequency of $\Gamma = 3\,{\rm hr}^{-1}$ for KMTC 
and $\Gamma = 2.25\,{\rm hr}^{-1}$ for KMTS and KMTA.

For both events, the data sets were reduced based on the image subtraction methodology 
\citep{tomaney96,alard98}, specifically \citet{albrow09} for KMTNet and \citet{wozniak00} for OGLE. 
The photometric error bars were then readjusted following the prescription presented in \citet{yee12}. 
We note that for the source color measurement, we additionally carried out pyDIA \citep{albrow17} 
reductions for a subset of the KMTNet data, which simultaneously returns the light curve and 
field-star photometry on the same system.

\section{Light Curve Analysis}

\begin{figure}[th]
\epsscale{0.9}
\plotone{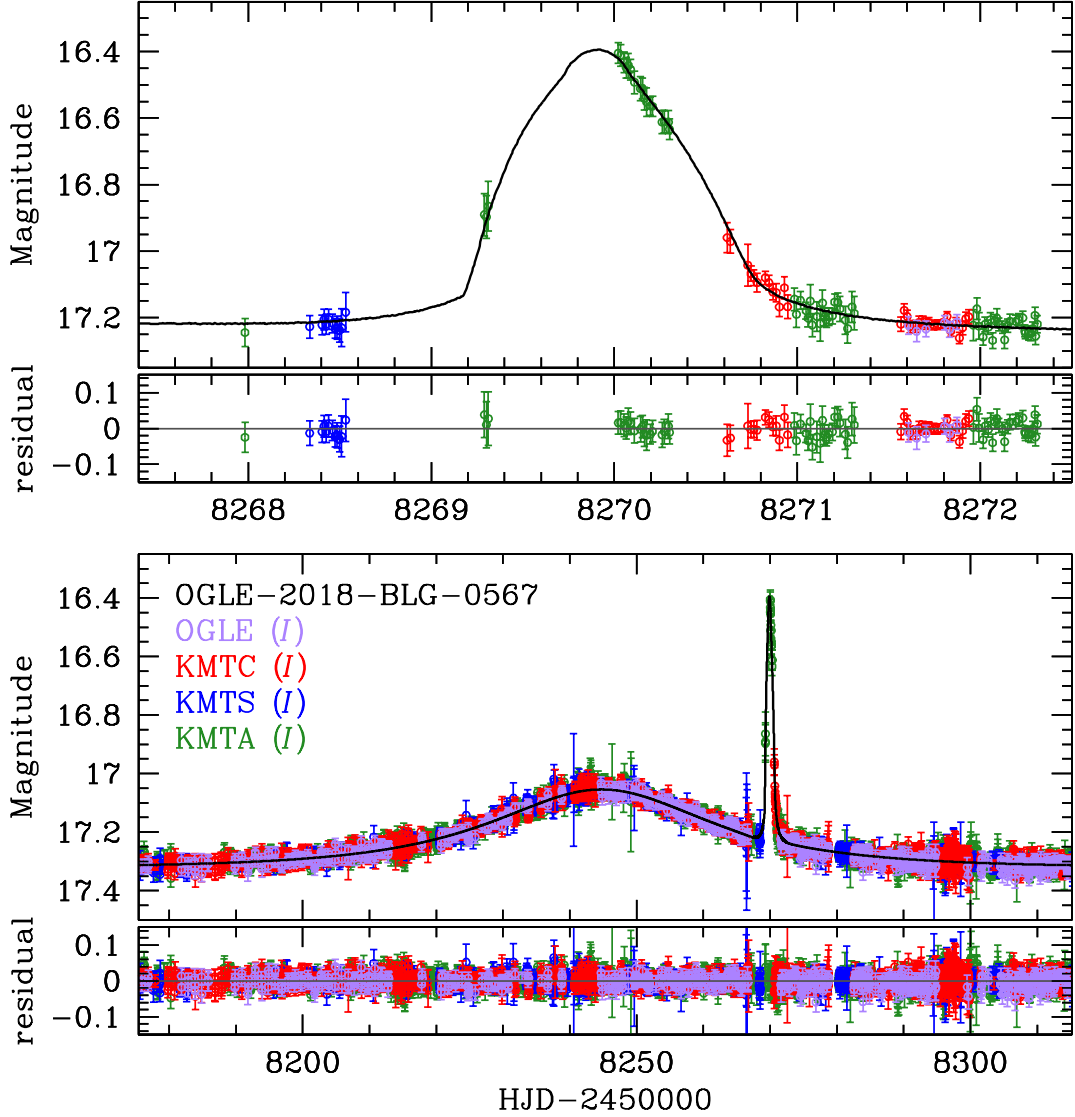}
\caption{\label{fig:one}
Light curve of OGLE-2018-BLG-0567.
The black solid curve on the data is the best-fit 2L1S solution. 
The upper panel shows the enlarged view of the planet-induced anomaly centered on ${\rm HJD}'\sim8270$.
The second and fourth panels show the residuals from the solution. 
The lensing parameters of the solution are listed in Table~\ref{table:2L1S} and the caustic geometry is shown in Figure~\ref{fig:three}.
}
\end{figure}

\begin{figure}[th]
\epsscale{0.9}
\plotone{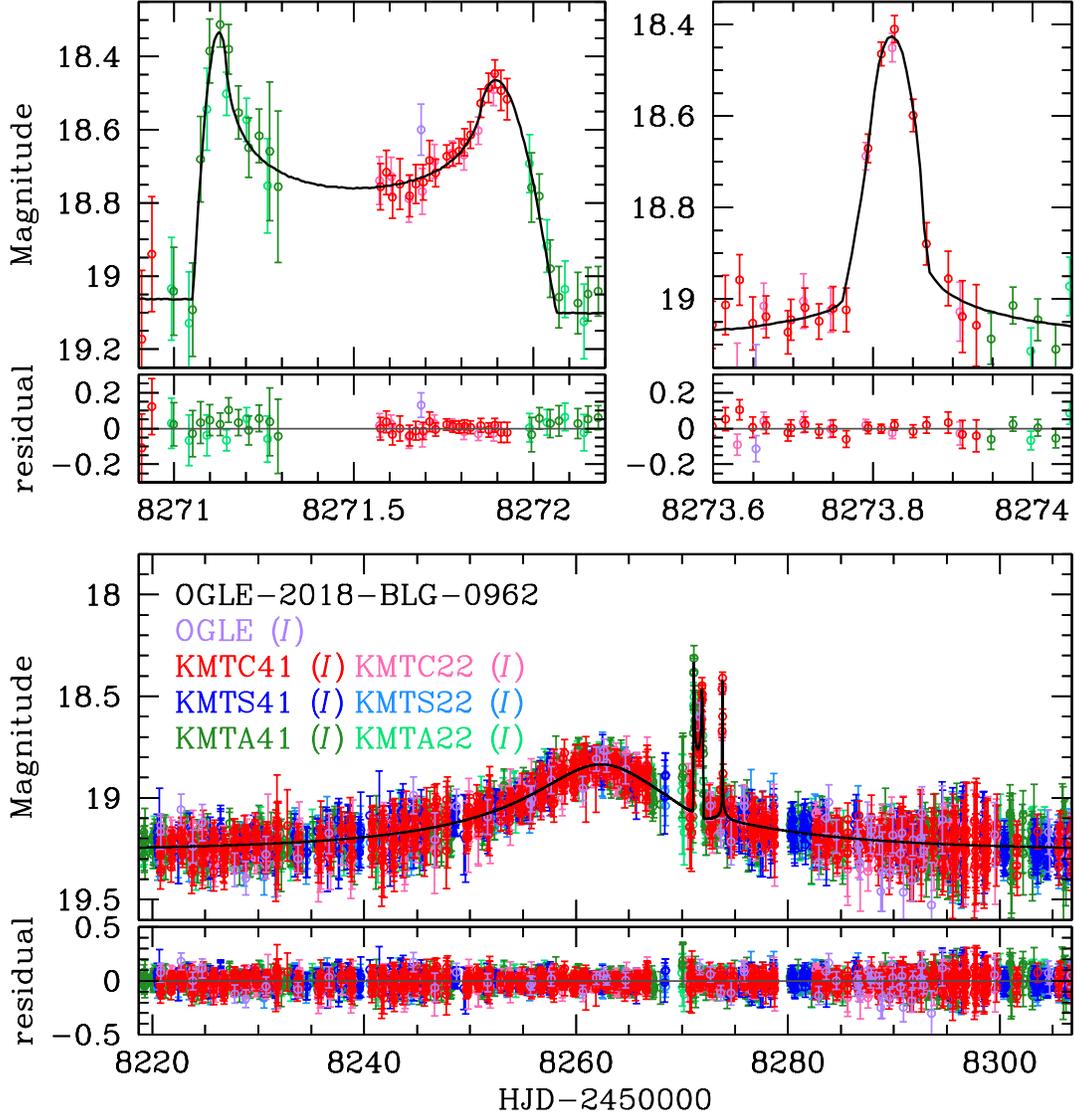}
\caption{\label{fig:two}
Light curve of OGLE-2018-BLG-0962. 
The upper panels show the close-up views of the regions around ${\rm HJD}'\sim8271.5$ (left) and ${\rm HJD}'\sim8273.8$ (right) when the planet-induced perturbations occur.
The lensing parameters of the 2L1S solution are listed in Table~\ref{table:2L1S} and the caustic geometry is shown in Figure~\ref{fig:four}.
}
\end{figure}

Figures~\ref{fig:one} and~\ref{fig:two} show the light curves of OGLE-2018-BLG-0567 and 
OGLE-2018-BLG-0962, respectively. It is found that the two events share various characteristics in common. 
First, the apparent peak magnifications of the baseline single-lens single-source (1L1S) 
light curves are not high: $A_{\rm peak} \sim 1.6$ for OGLE-2018-BLG-0567 and $A_{\rm peak} \sim 4.9$ 
for OGLE-2018-BLG-0962. Second, the light curves of both events exhibit strong short-term positive 
anomalies from the baseline 1L1S curves. Third, the anomalies appear when the 1L1S-model lensing 
magnifications are low. All these characteristics strongly suggest that the anomalies are 
produced by source crossings over the planetary caustics induced by planetary companions to 
the lenses. We, therefore, start with a binary-lens single-source (2L1S) modeling of the events 
under the interpretation that the events were produced by lenses composed of two masses, $M_1$ and $M_2$.

The standard 2L1S modeling requires one to include seven fitting parameters to describe 
an observed light curve. The first three are the \citet{paczynski86} parameters 
$(t_{0}, u_{0}, t_{\rm E})$, respectively the time of closest source approach to the lens, 
the impact parameter (scaled to $\theta_{\rm E}$), and the event timescale. 
The next three $(s, q, \alpha)$ describe the binary-lens geometry: 
the projected binary separation (scaled to $\theta_{\rm E}$), 
the binary mass ratio $(q = M_{2}/M_{1})$, and the orientation angle of 
the binary axis (relative to the source trajectory), respectively. 
The last describes the source radius $\rho = \theta_{*}/\theta_{\rm E}$, 
where $\theta_{*}$ is the angular source radius.

The modeling is conducted following the procedure described in \citet{jung15}. In the first step, 
we carry out grid searches for the binary parameters $(s, q, \alpha)$. At each grid, we fix $(s, q)$ 
and find the remaining parameters using Markov Chain Monte Carlo (MCMC) $\chi^{2}$ minimization. 
In this modeling, the initial values of the parameters $(t_{0}, u_{0}, t_{\rm E})$ are given as 
the values estimated from a 1L1S fit for the data excluding the anomaly. The initial value of 
the normalized source radius is estimated from the caustic-crossing timescale, which is related to 
the event timescale by $t_{*} = \rho t_{\rm E}$. Here we use inverse ray shooting  \citep{kayser86,schneider87} 
to compute the finite-source lensing magnifications. The flux values from the source, $f_{{\rm S},i}$, 
and blend, $f_{{\rm B},i}$, for the data set obtained from the $i$th observatory is estimated by 
$f_{i}(t) = f_{{\rm S},i}A(t) + f_{{\rm B},i}$, where $f_{i}$ is the observed flux. 
Once local solutions are found from the first-round modeling, we refine the individual locals by 
releasing all fitting parameters and allowing them to be free parameters in an MCMC.

From the modeling, it is found that the observed lensing light curves of both events are well 
described by unique 2L1S models, in which the mass ratios between $M_1$ and $M_2$ are in the 
planetary regime. The estimated binary parameters are $(s, q) = (1.81, 1.24\times10^{-3})$ 
for OGLE-2018-BLG-0567, and $(s, q) = (1.25, 2.38\times10^{-3})$ for OGLE-2018-BLG-0962. 
The full lensing parameters and their uncertainties are presented in Table~\ref{table:2L1S}. 
The model curves of the solutions are drawn over the data points in Figures~\ref{fig:one} 
and~\ref{fig:two} for OGLE-2018-BLG-0567 and OGLE-2018-BLG-BLG-0962, respectively.

\begin{figure}[th]
\epsscale{0.9}
\plotone{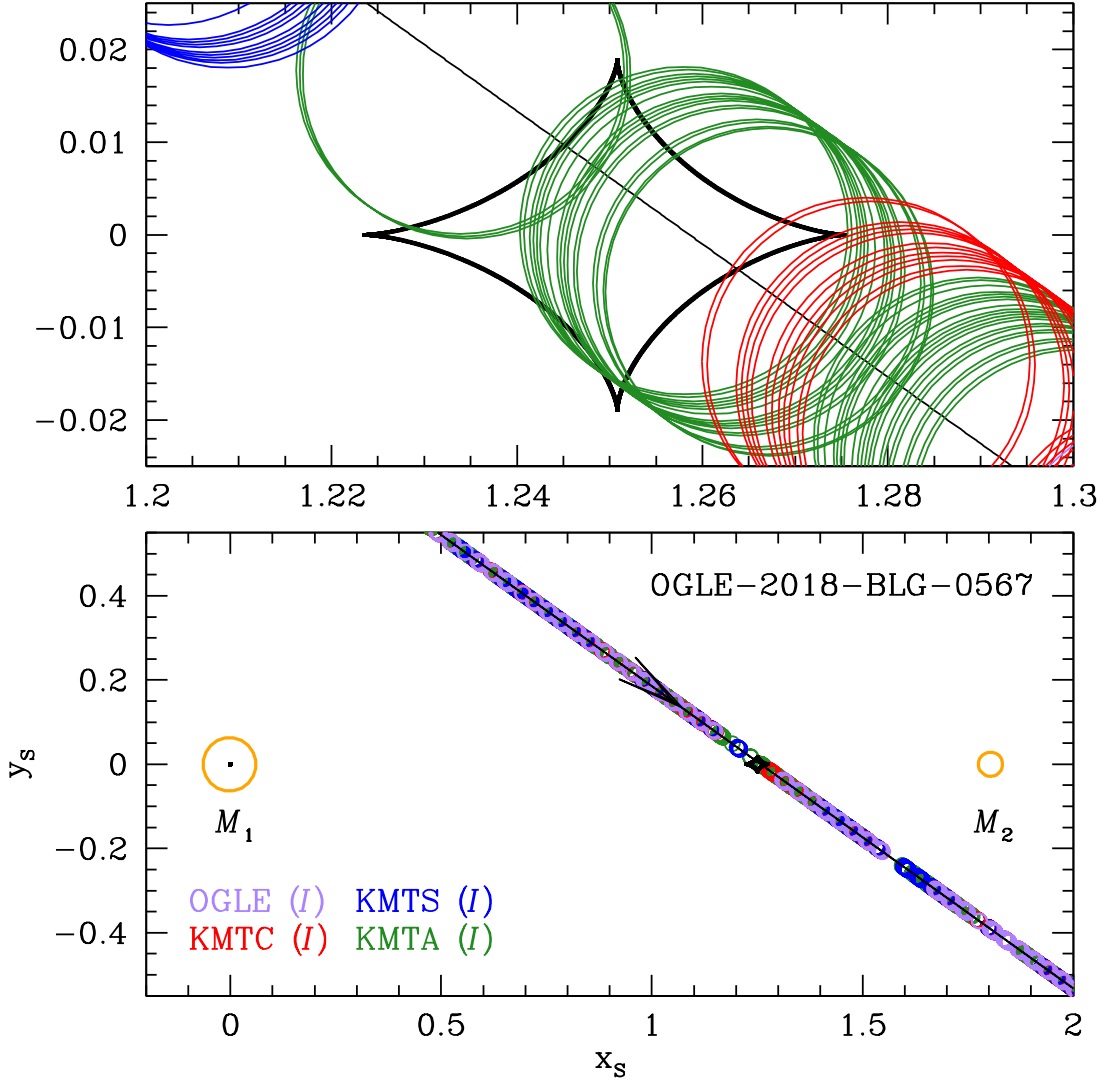}
\caption{\label{fig:three}
Caustic geometry of OGLE-2018-BLG-0567. 
The line with an arrow is the source trajectory relative to the binary axis. 
The open circles (scaled by the normalized source radius $\rho$) on the trajectory are the source positions at the times of observations. 
The two orange circles are the positions of binary-lens masses ($M_1$ and $M_2$).  
In each panel, the cuspy closed curve drawn in black color represents the caustic. 
The upper panel shows the enlarged view of the planetary caustic. 
Lengths are scaled to the angular Einstein radius of the lens system. 
}
\end{figure}

\begin{figure}[th]
\epsscale{0.9}
\plotone{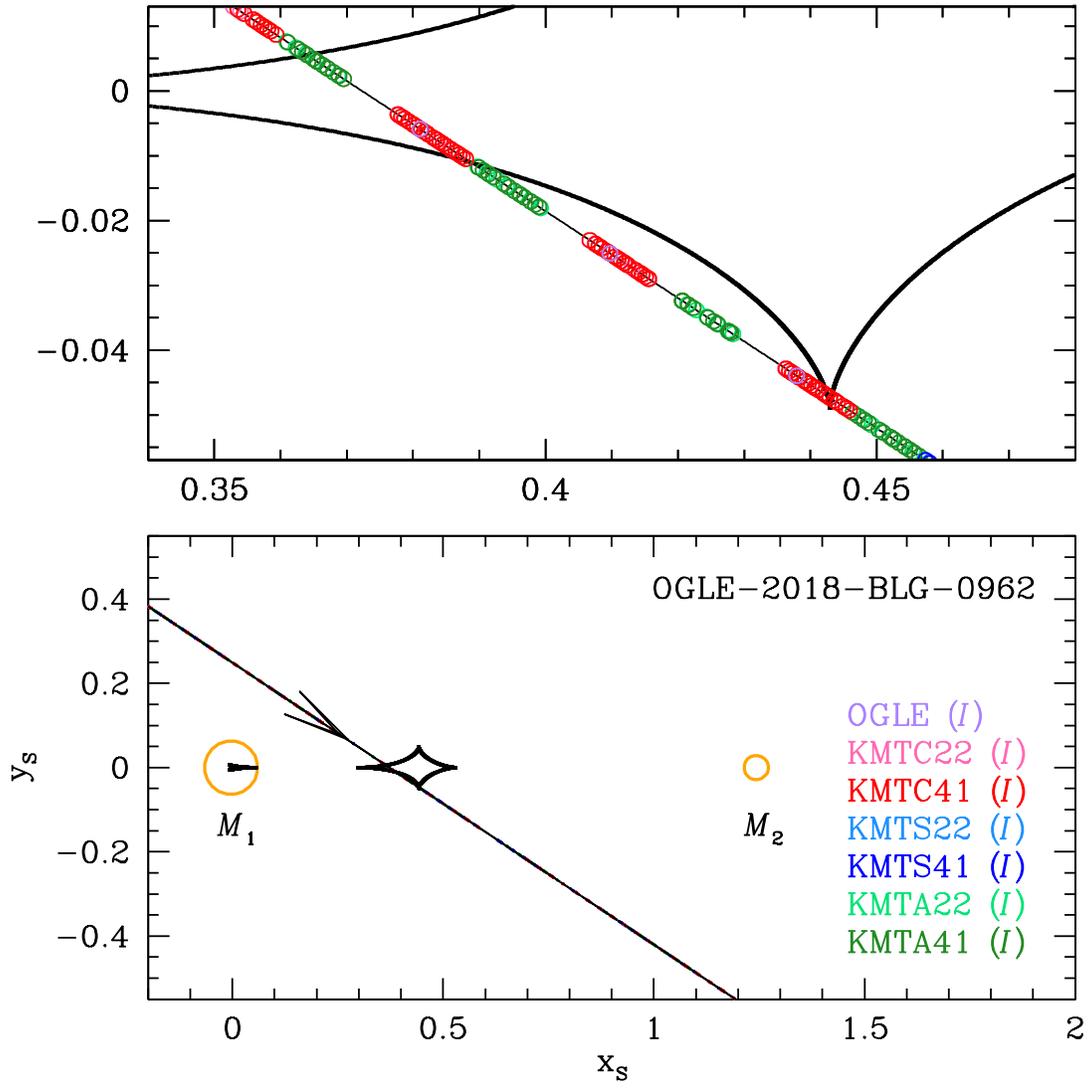}
\caption{\label{fig:four}
Caustic geometry of OGLE-2018-BLG-0962. 
Notations are identical to those of Figure~\ref{fig:three}.
}
\end{figure}

In Figures~\ref{fig:three} and~\ref{fig:four}, we present the lens-system configurations 
of the individual events, showing the source trajectory with respect to the lens components 
and resulting caustics. From the configurations, it is found that the anomalies of both events 
are produced by the source crossing over the planetary caustic of the lens system. 
For OGLE-2018-BLG-0567, the source size is comparable to the caustic size, and thus the detailed 
caustic-crossing features, two caustic spikes and a U-shape trough region between the spikes, 
was smeared out by finite-source effects. For OGLE-2018-BLG-0962, on the other hand, the caustic 
is much bigger than the source size, and thus detailed caustic-crossing feature of the anomaly 
is well delineated. It is found the first part of the anomaly, centered at 
${\rm HJD}'(={\rm HJD}-2,450,000\,{\rm days}) \sim 8271.5$, was produced by the source passing 
over the two caustic segments that flank the inner cusp (on the binary axis) of the planetary caustic, 
and the second part, centered at ${\rm HJD}' \sim 8273.8$, was generated by the 
source passage over the adjacent (off-axis) cusp.

\begin{figure}[th]
\epsscale{0.9}
\plotone{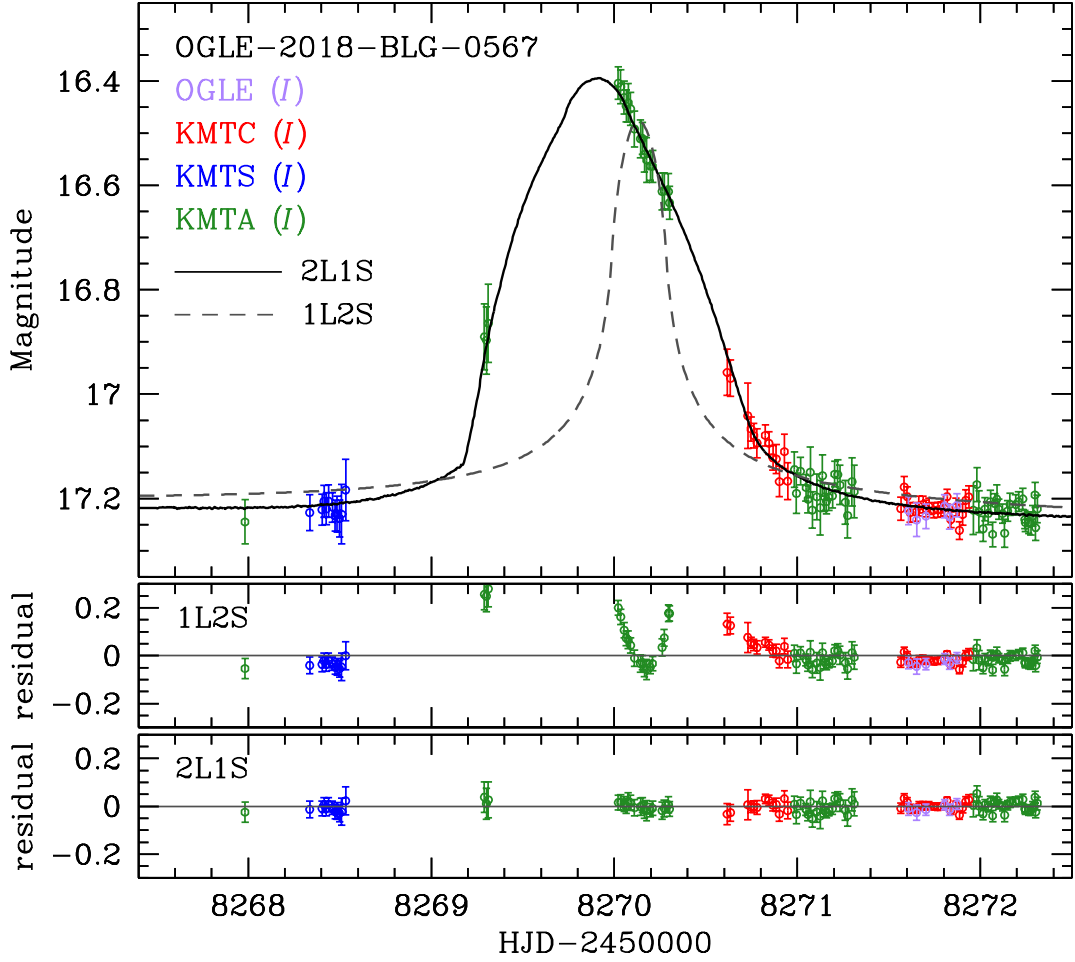}
\caption{\label{fig:five}
Light curve of the 1L2S model for OGLE-2018-BLG-0567. 
The dashed gray and solid black lines are the best-fit models from the 1L2S and 2L1S interpretations, respectively. 
The lower two panels show the residuals from the two models.
}
\end{figure}

We investigate the possibility of other interpretations of the events. 
Especially for OGLE-2018-BLG-0567, the short-term perturbation might, in principle, be produced 
by a second source (1L2S) with a large flux ratio \citep{gaudi98}. Hence, we conduct an additional 
modeling of the event with the 1L2S interpretation \citep{jung17}. We search for the solution with eight 
fitting parameters: $2\times(t_{0}, u_{0}, \rho)$ for the two sources, $I$-band flux ratio $q_{F,I}$, 
and a shared timescale $t_{\rm E}$. The results are listed in Table~\ref{table:1L2S}. We find that 
the 1L2S solution is disfavored by $\Delta\chi^{2} > 500$. In addition, the 1L2S model not only 
provides a worse fit to the peak of the perturbation, but also fails to recover the decrease 
in magnitude seen before and after the peak. See Figure~\ref{fig:five}. We also investigate 
higher-order effects by conducting additional modeling considering the microlens-parallax and 
lens-orbital effects \citep{gould92a,dominik98}, and find that it is difficult to constrain 
the extra lensing parameters related to these higher-order effects from the observed light curves.

\section{Physical Parameters}

The results in Table~\ref{table:2L1S} show that for both events, the normalized source radii are 
precisely measured. This enables us to estimate $\theta_{\rm E} = \theta_{*}/\rho$ provided that 
$\theta_{*}$ is measured. Then, we can use the estimated $\theta_{\rm E}$ to constrain the lens 
total mass $M$ and distance $D_{\rm L}$ as given by
\begin{equation}
\theta_{\rm E}^{2} \equiv \kappa M \pi_{\rm rel};
\qquad
\pi_{\rm rel} = {\rm au}\left({{1 \over D_{\rm L}} - {1 \over D_{\rm S}}}\right),
\label{thetae}
\end{equation}
where $\kappa\equiv {4G/(c^2{\rm au})} \simeq 8.14\,{{\rm mas}/M_{\odot}}$, 
$\pi_{\rm rel}$ is the lens-source relative parallax, and $D_{\rm S}$ is the source distance. 
Hence, we evaluate $\theta_{*}$ using the method of \citet{yoo04}.

\begin{figure}[th]
\epsscale{0.9}
\plotone{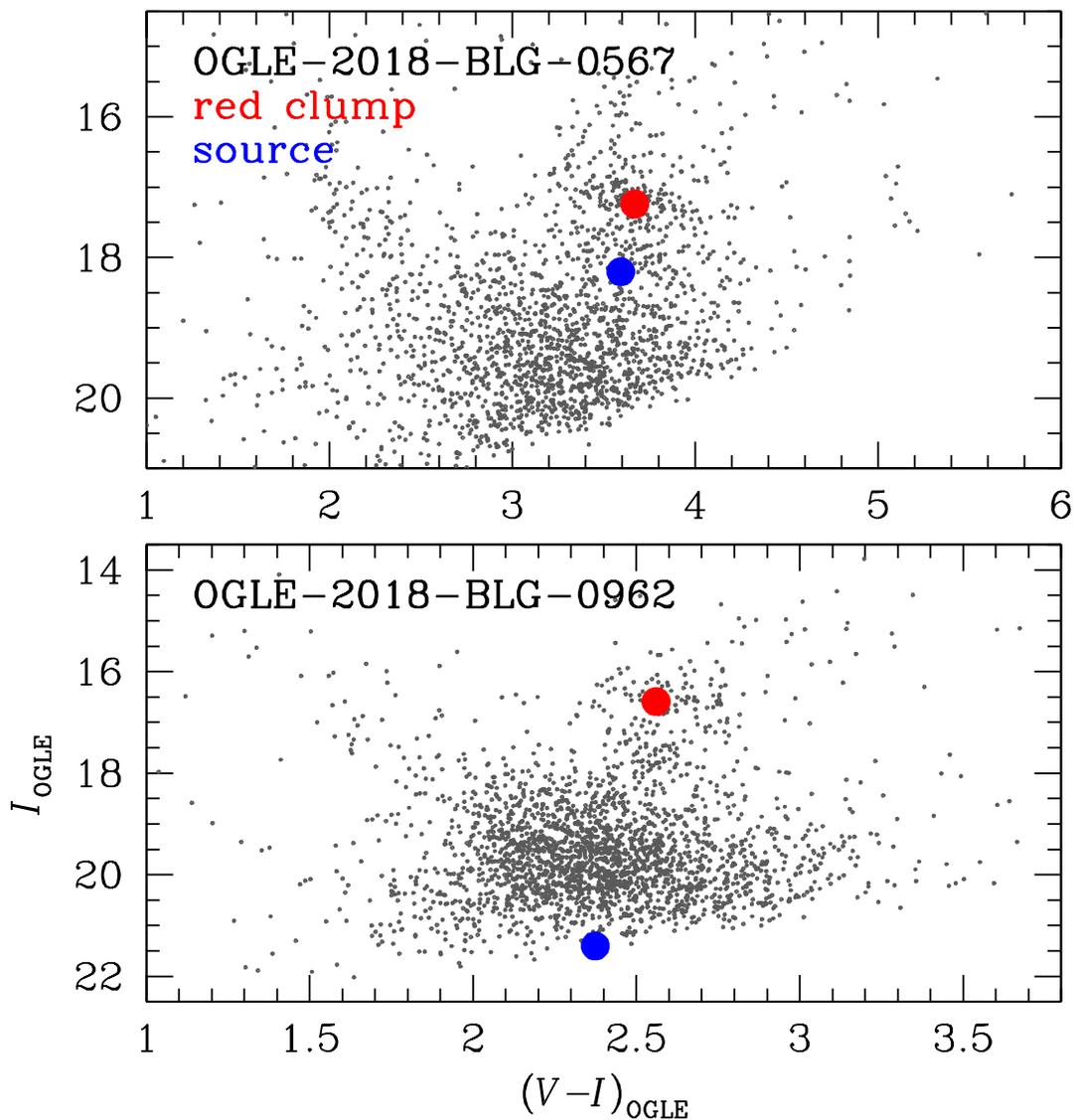}
\caption{\label{fig:six}
Color-magnitude diagrams of OGLE-2018-BLG-0567 (upper panel) and OGLE-2018-BLG-0962 (lower panel). 
In each panel, the CMD is constructed using stars in the $2'\times2'$ field centered on the event 
location based on KMTNet pyDIA photometry calibrated to the OGLE-III catalog \citep{szymanski11}.
The blue and red circles are the positions of source and red clump centroid, respectively.
}
\end{figure}

We first estimate the intrinsic source color $(V-I)_{\rm 0,S}$ and magnitude $I_{\rm 0,S}$. 
We do this in the following ways. First, we calibrate the KMTNet pyDIA photometry to the 
standard Johnson-Cousins system using the OGLE-III catalog \citep{szymanski11}. We next 
construct a $(V-I, I)$ color-magnitude diagram (CMD) with field stars around the source. 
We next find the source location $(V-I, I)_{\rm S}$ in the CMD from the best-fit model. 
We also find the position of the red clump centroid (RCC), i.e., $(V-I, I)_{\rm RCC}$. 
Figure~\ref{fig:six} shows the source and RCC positions in the CMDs for the individual events. 
We then measure the offset $\Delta(V-I, I) = (V-I, I)_{\rm S} - (V-I, I)_{\rm RCC}$. 
Finally, we find the intrinsic source position as
\begin{equation}
(V-I, I)_{\rm 0,S} = \Delta(V-I, I) + (V-I, I)_{\rm 0,RCC}, 
\label{source}
\end{equation}
where $(V-I, I)_{\rm 0,RCC}$ is the intrinsic RCC position measured from independent 
observations \citep{bensby13,nataf13}. Here we assume that the source star experiences 
the same amount of extinction as the RCC. In Table~\ref{table:source}, we list our estimated 
values of $(V-I, I)_{\rm S}$, $(V-I, I)_{\rm RCC}$, $(V-I, I)_{\rm 0,RCC}$, and $(V-I, I)_{\rm 0,S}$.

We now derive $\theta_{*}$ based on the estimated intrinsic source position. For this, 
we apply $(V-I)_{\rm 0,S}$ to the $VIK$ relation \citep{bessell88} to find $(V-K)_{\rm 0,S}$. 
We then obtain $\theta_{*}$ from the $(V-K)_{\rm 0,S}-\theta_{*}$ relations, specifically 
\citet{kervella04a} for OGLE-2018-BLG-0567 and \citet{kervella04b} for OGLE-2018-BLG-0962. 
We note that we add $5\%$ error to $\theta_{*}$ to consider the uncertainty of the intrinsic 
RCC position and the color/surface brightness conversion. From the measured $t_{\rm E}$ and $\rho$, 
we then find $\theta_{\rm E}$ and the lens-source relative proper motion, 
$\mu_{\rm rel} = \theta_{\rm E}/t_{\rm E}$. The estimated values of $\theta_{*}$, $\theta_{\rm E}$, 
and $\mu_{\rm rel}$ are also listed in Table~\ref{table:source}.

For both events, we are unable to constrain the microlens parallax vector $\pivec_{\rm E}$. 
This implies that we cannot directly derive $M$ and $D_{\rm L}$ by
\begin{equation}
M = {\theta_{\rm E} \over \kappa\pi_{\rm E}};
\qquad
D_{\rm L} = {{\rm au} \over \pi_{\rm E}\theta_{\rm E} + \pi_{\rm S}}
\label{properties}
\end{equation}
from the microlensing data \citep{gould00}. Here $\pi_{\rm S} = {\rm au}/D_{\rm S}$. 
Hence, we estimate the lens properties based on a Bayesian analysis 
with Galactic model priors.

The Galactic model is constructed based on a mass function (MF), a density profile (DP), 
and a velocity distribution (VD). For the MF, we use the models presented in \citet{jung18}. 
For the DP, we use the \citet{han03} model for the bulge and the \citet{robin03} model for the disk, 
respectively. We note that the former is normalized based on the star counts results of \citet{holtzman98}, 
while the latter is normalized by the local column density 
$\Sigma(D_{\rm L}=0) = \Sigma_{\rm disk,0} = 36\,{M_\odot}\,{{\rm pc}^{-2}}$ from \citet{han03}.

The bulge VD is modeled based on stars in the Gaia catalog \citep{gaia16,gaia18}. That is, 
we first find red giant stars in the catalog within 2 arcmin centered on the event location. 
We then derive their mean velocity and its dispersion in Galactocentric Cartesian coordinates 
$(x,y,z)$ as defined in \citet{han95}.

For the disk VD, we adopt Gaussian forms of $f(v_{y}, v_{z}) = f(v_{y})f(v_{z})$ from \citet{han95}, 
which we then modify to consider the change in the matter distribution. We do this in two ways. 
First, we introduce an asymmetric drift $v_{\rm ad}$ to the $y$-direction velocity $v_y$. 
That is, $v_{\rm ad}(D_{\rm L}) = (0.5\sigma_{xyz,0}^{2}/\bar{v}_{y})[\Sigma(D_{\rm L})/\Sigma_{\rm disk,0}]^{1/2}$, 
where $\bar{v}_{y} = 220\,{\rm km}\,{\rm s}^{-1}$ is the mean $y$-direction velocity, 
$\sigma_{xyz,0}^{2} = \sigma_{x,0}^2 + \sigma_{y,0}^2 + \sigma_{z,0}^2$, and 
$(\sigma_{x,0}, \sigma_{y,0}, \sigma_{z,0}) = (34, 30, 20)\,{\rm km}\,{\rm s}^{-1}$ 
are the mean velocity dispersions along the $(x, y, z)$ directions (in the solar neighborhood), respectively. 
The $y$-direction velocity at a given line-of-sight distance $D_{\rm L}$ is then calculated by 
$v_{y}(D_{\rm L}) = \bar{v}_{y} - v_{\rm ad}(D_{\rm L})$. 
Second, we subsequently modify the velocity dispersion as 
$\sigma_{y}(D_{\rm L}) = \sigma_{y,0}[\Sigma(D_{\rm L})/\Sigma_{\rm disk,0}]^{1/2}$ for the $y$-direction and 
$\sigma_{z}(D_{\rm L}) = \sigma_{z,0}[\Sigma(D_{\rm L})/\Sigma_{\rm disk,0}]^{1/2}$ for the $z$-direction.

We now carry out the Bayesian analysis with the constraints $(t_{\rm E}, \theta_{\rm E})$. 
For this, we follow the procedure of \citet{jung18}. The estimated lens properties for the 
individual events are listed in Table~\ref{table:phys}. 
The corresponding posterior probabilities for $M_{1}$ and $D_{\rm L}$ are shown in 
Figure~\ref{fig:seven}. We find that the host mass is  
$M_{1} = 0.24_{-0.13}^{+0.26}\,M_{\odot}$ for OGLE-2018-BLG-0567 and
$M_{1} = 0.55_{-0.29}^{+0.32}\,M_{\odot}$ for OGLE-2018-BLG-0962. 
The planet masses $(M_{2}=qM_{1})$ and the projected planet-host separations 
$(a_{\perp}=sD_{\rm L}\theta_{\rm E})$ of the individual events are then estimated to 
$(M_{2}, a_{\perp}) = (0.32_{-0.16}^{+0.34}\,M_{\rm J}, 2.72_{-0.55}^{+0.49}\,{\rm au})$ and 
$(M_{2}, a_{\perp}) = (1.37_{-0.72}^{+0.80}\,M_{\rm J}, 3.57_{-1.02}^{+0.68}\,{\rm au})$, respectively. 
These estimates suggest that the two planetary systems are likely 
composed of M-type dwarfs and giant planets lying beyond the snow line \citep{kennedy08}. 
Here, the snow line is the location in the proto-planetary disk where icy material can condense 
and where giant planets are thought to be formed \citep{ida04}. The distance to the planet is  
$D_{\rm L} = 7.07_{-1.15}^{+0.93}\,{\rm kpc}$ for OGLE-2018-BLG-0567 and 
$D_{\rm L} = 6.47_{-1.73}^{+1.04}\,{\rm kpc}$ for OGLE-2018-BLG-0962, 
indicating that they are likely to be in or near the Galactic bulge.

\section{Microlensing Planets in the $({\rm log}\,s, {\rm log}\,q)$ plane}

Our two survey-only microlensing planets are detected from the perturbations 
caused by the planetary caustics (see Figures~\ref{fig:three} and~\ref{fig:four}). 
In particular, the planetary perturbation of OGLE-2018-BLG-0567 was generated by 
a ``Hollywood'' geometry \citep{gould97a}, in which the source size contributes 
strongly to, or dominates, the anomaly cross section relative to the size of the caustic. 
These detections prove the capability of the high-cadence surveys for detecting planets 
through the planetary-caustic channel.

Figure~\ref{fig:eight} illustrates the positions of OGLE-2018-BLG-0567Lb and OGLE-2018-BLG-0962Lb 
in the $({\rm log}\,s, {\rm log}\,q)$ plane along with other microlensing planets\footnote{https://exoplanetarchive.ipac.caltech.edu}. 
This figure shows that OGLE-2018-BLG-0567Lb is located in a previously underpopulated region. 
The two green solid lines indicate the boundary between resonant and non-resonant caustics \citep{schneider86,dominik99}. 
For planets just outside of the resonance region, the excess magnification pattern 
(from the underlying 1L1S magnification) extends out all the way between the central and planetary caustics. 
This implies that the effective cross-section of these caustics for a planet-induced perturbation and their 
lensing behavior are similar to those of resonant caustics. 
Recently, \citet{yee21} classified such kinds of caustics as the ``near-resonant'', 
specifically a set of caustics that has a excess magnification contour that connects the central and 
planetary caustics and has at least $10\%$ excess magnification along the entire caustic ridges. 
The two green dashed lines are the boundary of near-resonant caustics. 
\citet{yee21} empirically estimated that at a fixed $q$, the maximum size of a $10\%$ deviation 
corresponds to $\sim 3{\rm log}\,{s_{\rm r,c}}$ for $s < 1$ (close) and 
$\sim 1.8{\rm log}\,{s_{\rm r,w}}$ for $s >1$ (wide), where $s_{\rm r,c}$ and $s_{\rm r,w}$ are the 
boundary values of $s$ between resonant and non-resonant caustics \citep{dominik99}. 
Planets (except for our two planets) with a single and multiple solutions are coded by black 
and red (with connected lines) colors, respectively. OGLE-2018-BLG-0567Lb and OGLE-2018-BLG-0962Lb 
are marked by yellow and blue colors, respectively. The shape of the symbols represent the type of 
caustics that yield the planetary perturbation: circles for resonant/near-resonant, 
squares for central, and triangles for planetary caustics. 
The filled triangles are the planets found in the Hollywood events.

Figure 8 shows that the majority of planets are located inside the near-resonant boundary rather than 
being due to planetary caustics. The bias toward resonant caustics mainly comes from the relatively large 
size of resonant caustics (scaled as $q^{1/3}$) compared to that of planetary caustics (scaled as $q^{1/2}$) 
\citep{dominik99,han06,yee21}. There is also a bias against planetary caustics due to the observational 
strategy of earlier microlensing experiments that was focused on high-magnification events \citep{griest98,gould10}.

Only $24$ planets are placed outside the near-resonant boundary and $18$ planets among them 
are detected from the perturbations produced by clearly isolated planetary caustics\footnote{
The corresponding planetary-caustic events are 
OGLE-2005-BLG-390 \citep{beaulieu06}, 
MOA-bin-1 \citep{bennett12},
OGLE-2006-BLG-109 \citep{gaudi08,bennett10}, 
OGLE-2008-BLG-092 \citep{poleski14}, 
MOA-2010-BLG-353 \citep{rattenbury15}, 
MOA-2011-BLG-028 \citep{skowron16}, 
MOA-2012-BLG-006 \citep{poleski17}, 
OGLE-2012-BLG-0838 \citep{poleski20}, 
OGLE-2013-BLG-0341 \citep{gould14}, 
MOA-2013-BLG-605 \citep{sumi16}, 
OGLE-2014-BLG-1722 \citep{suzuki18}, 
OGLE-2016-BLG-0263 \citep{han17}, 
OGLE-2016-BLG-1227 \citep{han20}, 
KMT-2016-BLG-1107 \citep{hwang19}, 
OGLE-2017-BLG-0173 \citep{hwang18},
OGLE-2017-BLG-0373 \citep{skowron18}, 
OGLE-2018-BLG-0596 \citep{jung19}, 
and OGLE-2018-BLG-0962 (this work).}. 
We find that most of these planetary-caustic planets ($12$ planets) are found in the Hollywood events 
and they are located in high-cadence observational fields of the lensing surveys. This proves the capability 
of the Hollywood strategy of following big stars to find planets \citep{gould97a}. The majority of the 
Hollywood planets are located in the region $s>1$. This is mainly due to the difference in the size of 
planetary caustics. For $s > 1$, there is one four-sided planetary caustic. For $s < 1$, on the other hand, 
there are two triangular planetary caustics and each of which size is much smaller than that of $s > 1$. 
In addition, the planetary signals from the these smaller planetary caustics tends to be more significantly 
diminished by the finite-source effects \citep{gould97b}. As a result, the wide-planetary caustic 
has a larger effective cross section and therefore higher sensitivity for finding planets.

\section{Summary and Conclusions}

We present the discovery of two cold giant planets orbiting M-dwarfs in two events OGLE-2018-BLG-0567 and OGLE-2018-BLG-0962. 
Both events clearly showed deviations from the 1L1S model, caused by the presence of a companion to the lens host with precisely 
measured planet/host mass ratios of $(1.24\pm0.07)\times10^{-3}$ and $(2.38\pm0.08)\times10^{-3}$, respectively. In both events, 
the finite-source effects are clearly detected, but the microlens parallax effects are not meaningfully constrained. 
Hence, we constrain the lens properties using the Bayesian analysis. From this, we estimate planet masses of 
$0.32_{-0.16}^{+0.34}\,M_{\rm J}$ for OGLE-2018-BLG-0567Lb and $1.37_{-0.72}^{+0.80}\,M_{\rm J}$ for OGLE-2018-BLG-0962Lb, 
and their physical projected separations of $2.72_{-0.55}^{+0.49}\,{\rm au}$ and $3.57_{-1.02}^{+0.68}\,{\rm au}$, respectively. 
These planets likely belong to a class of giant planets orbiting M-dwarfs outside the snow line. 
The detection rate of microlensing planets has rapidly increased with the advent of high-cadence lensing surveys, and thus, 
planets presented here and future detections will expand our understanding of the planet population around M-dwarfs.

\begin{figure}[th]
\epsscale{0.9}
\plotone{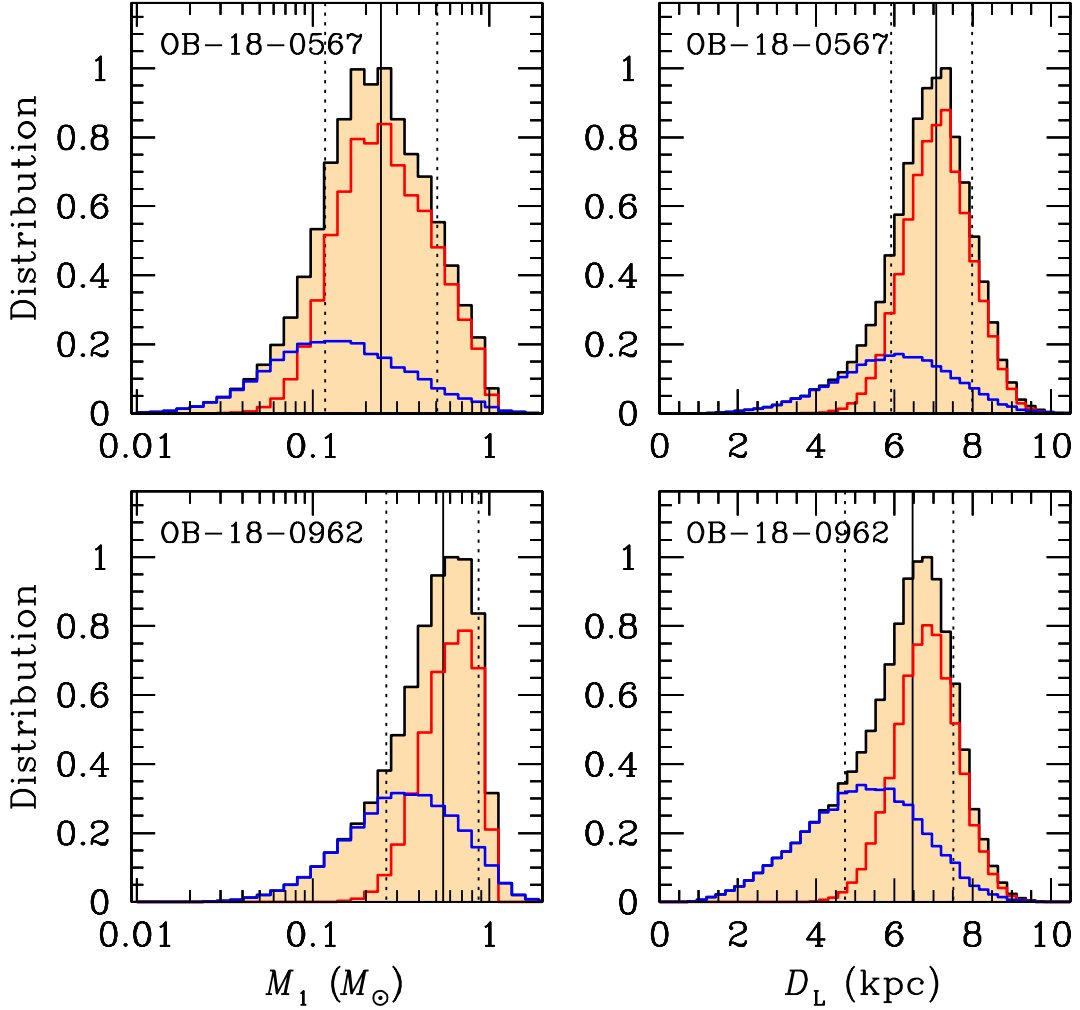}
\caption{\label{fig:seven}
Posterior distributions of $M_{1}$ (left panels) and $D_{\rm L}$ (right panels) for the individual events.
In each panel, the red and blue distributions are, respectively, the contributions by the bulge and disk lens populations. 
The black distribution is the total contribution of the two lens populations. 
The median value and its $68\%$ confidence interval are represented by the vertical solid and two dotted lines, respectively.
}
\end{figure}

\begin{figure}[th]
\epsscale{1.0}
\plotone{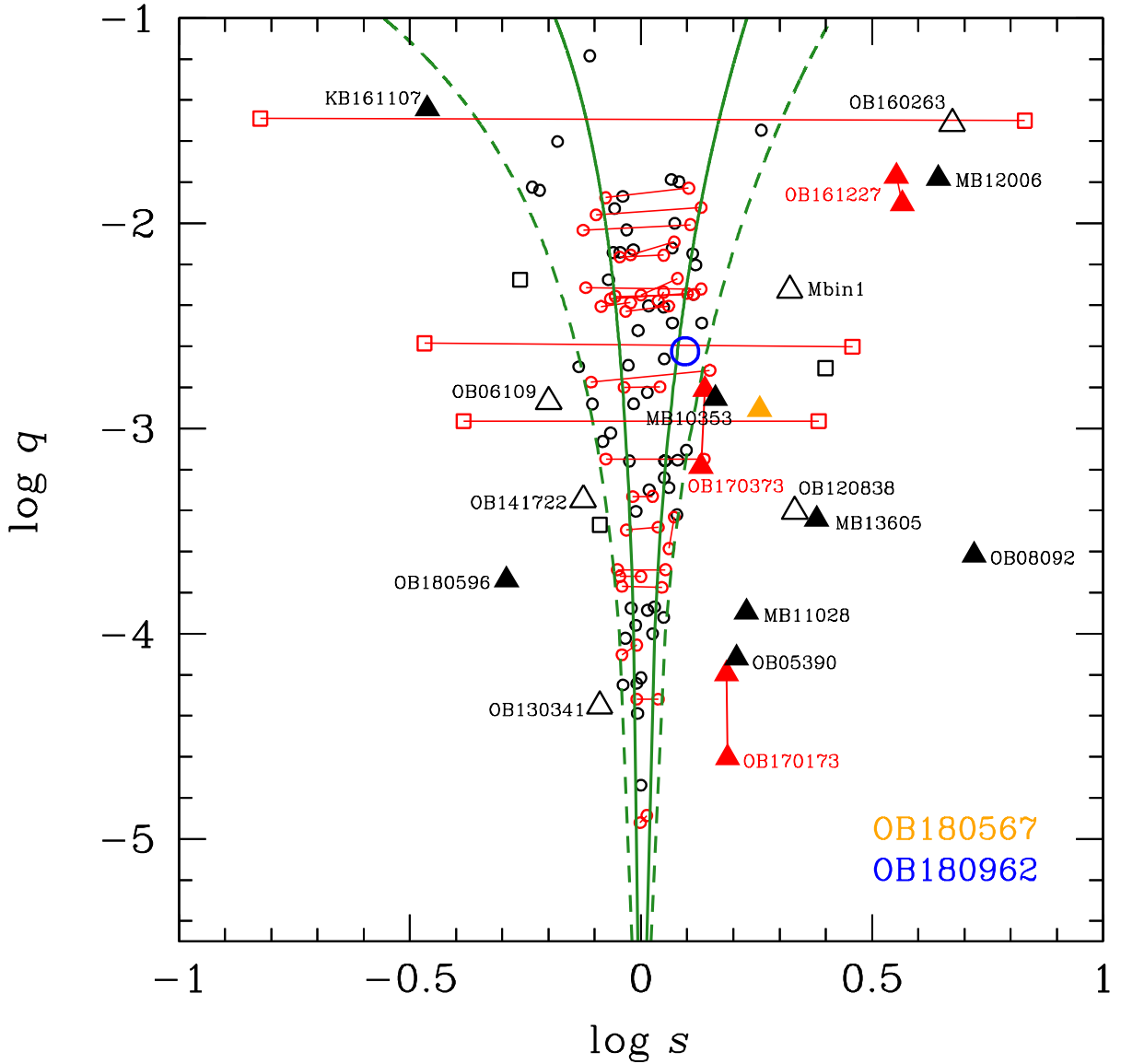}
\caption{\label{fig:eight}
Microlensing planets in the $({\rm log}\,s, {\rm log}\,q)$ plane, adapted from Figure~9 of \citet{yee21}. 
Planets (except for our two planets) are colored by the number of solutions: black for one solution and red 
(with connected line) for degenerate solutions. The two planets OGLE-BLG-2018-BLG-0567Lb and OGLE-2018-BLG-0962Lb 
are coded by yellow and blue colors, respectively. Their shapes indicate the caustic structure giving rise to the 
planetary perturbation: circles for resonant/near-resonant, squares for central, and triangles for planetary caustics. 
The filled triangles are the planets from the Hollywood events. 
The two green solid and dashed lines are the boundary of resonant and near-resonant caustics, respectively. 
We note that for compactness, we compress the planet names, e.g., OGLE-2018-BLG-0567Lb to OB180567. 
}
\end{figure}

The planet hosts can be precisely constrained by future high-resolution imaging with adaptive optics (AO) mounted 
on ``$30\,{\rm m}$'' class telescopes. That is, they have only a small probability of being non-luminous 
(see Figure~\ref{fig:seven}). The Bayesian estimates suggest that the dereddened $H$-band magnitude of the host is 
$H_{0} = 21.90_{-1.82}^{+1.86}$ for OGLE-2018-BLG-0567 and $H_{0} = 19.47_{-1.52}^{+2.04}$ for OGLE-2018-BLG-0962 \citep{pecaut13}. 
In addition, both events have $\mu_{\rm rel} > 3.0\,{\rm mas}\,{\rm yr}^{-1}$. Therefore, in 2030, the hosts 
will be separated from the microlensed source by $\Delta\theta \gtrsim 40\,{\rm mas}$.

\acknowledgments
This research has made use of the KMTNet system operated by the Korea 
Astronomy and Space Science Institute (KASI) and the data were obtained at 
three host sites of CTIO in Chile, SAAO in South Africa, and SSO in Australia. 
Work by CH was supported by the grants of National Research Foundation of Korea (2017R1A4A1015178 and 2019R1A2C2085965).
The OGLE has received funding from the National Science Centre, Poland, grant MAESTRO 2014/14/A/ST9/00121 to A.U.

\begin{deluxetable}{lrrrr}
\tablecaption{Lensing Parameters}
\tablewidth{0pt}
\tablehead{
\multicolumn{1}{l}{Parameters} &
\multicolumn{1}{c}{OGLE-2018-BLG-0567} &
\multicolumn{1}{c}{OGLE-2018-BLG-0962} 
}
\startdata
$\chi^2_{\rm tot}$/dof          &     8677.3/9252        &    6892.1/6833       \\
$t_0$ (${\rm HJD'}$)            & 8244.845$\pm$ 0.025    & 8262.494$\pm$ 0.053    \\
$u_0$                           &   0.733 $\pm$ 0.026    &   0.207 $\pm$ 0.032    \\
$t_{\rm E}$ (days)              &  24.641 $\pm$ 1.064    &  28.739 $\pm$ 0.298    \\
$s$                             &   1.806 $\pm$ 0.019    &   1.246 $\pm$ 0.027    \\
$q$ ($10^{-3}$)                 &   1.240 $\pm$ 0.068    &   2.375 $\pm$ 0.076    \\
$\alpha$ (rad)                  &   0.623 $\pm$ 0.045    &   0.590 $\pm$ 0.044    \\
$\rho$ ($10^{-3}$)              &  17.675 $\pm$ 0.831    &   1.137 $\pm$ 0.048    \\
$f_{\rm S}$                     &   0.842 $\pm$ 0.035    &   0.039 $\pm$ 0.007    \\  
$f_{\rm B}$                     &   1.026 $\pm$ 0.035    &   0.274 $\pm$ 0.007    
\enddata 
\vspace{0.05cm}
\label{table:2L1S}
\end{deluxetable}

\begin{deluxetable}{lr}
\tablecaption{Binary Source Model for OGLE-2018-BLG-0567}
\tablewidth{0pt}
\tablehead{
\multicolumn{1}{l}{Parameters} &
\multicolumn{1}{c}{1L2S}
}

\startdata
$\chi^2$/dof                     &     9223.0/9251      \\
$t_{0,1}$ (HJD$'$)               &  8244.323$\pm$0.043  \\
$u_{0,1}$                        &     0.705$\pm$0.039  \\
$t_{0,2}$ (HJD$'$)               &  8670.134$\pm$0.010  \\
$u_{0,2}$ $(10^{-5})$            &     2.603$\pm$35.093 \\
$t_{\rm E}$ (days)               &    24.088$\pm$0.868  \\
$\rho_{1}$                       &     --               \\
$\rho_{2}$ $(10^{-3})$           &     6.295$\pm$0.344  \\
$q_{F,I}$ $(10^{-3})$            &     7.520$\pm$0.461  \\
$f_{{\rm s}}$                    &     0.795$\pm$0.082  \\
$f_{{\rm b}}$                    &     1.073$\pm$0.082   
\enddata  
\vspace{0.05cm}
\label{table:1L2S}
\end{deluxetable}

\begin{deluxetable}{lrr}
\tablecaption{Source Star and Lens Properties}
\tablewidth{0pt}
\tablehead{
\multicolumn{1}{l}{Parameters} &
\multicolumn{1}{c}{OGLE-2018-BLG-0567} &
\multicolumn{1}{c}{OGLE-2018-BLG-0962} 
}
\startdata
$(V-I, I)_{\rm S}$                            &    (3.59$\pm$0.10, 18.20$\pm$0.01)   &    (2.37$\pm$0.08, 21.40$\pm$0.03) \\
$(V-I, I)_{\rm RCC}$                          &    (3.67$\pm$0.02, 17.24$\pm$0.02)   &    (2.56$\pm$0.02, 16.59$\pm$0.03) \\
$(V-I, I)_{0,\rm RCC}$                        &    (1.06, 14.37)                     &    (1.06, 14.56)                   \\
$(V-I, I)_{0,\rm S}$                          &    (0.98$\pm$0.10, 15.33$\pm$0.03)   &    (0.87$\pm$0.08, 19.37$\pm$0.04) \\
$\theta_{*}$ $({\mu}{\rm as})$                &    3.769$\pm$0.419                   &    0.504$\pm$0.048                 \\
$\theta_{\rm E}$ (mas)                        &    0.213$\pm$0.026                   &    0.443$\pm$0.045                 \\     
$\mu_{\rm rel}$ $({\rm mas}~{\rm yr}^{-1})$   &    3.161$\pm$0.381                   &    5.639$\pm$0.572                    
\enddata 
\vspace{0.05cm}
\label{table:source}
\end{deluxetable}

\begin{deluxetable}{lrr}
\tablecaption{Physical Parameters}
\tablewidth{0pt}
\tablehead{
\multicolumn{1}{l}{Parameters} &
\multicolumn{1}{c}{OGLE-2018-BLG-0567} &
\multicolumn{1}{c}{OGLE-2018-BLG-0962} 
}
\startdata
$M_{1}$ $(M_{\odot})$             &    $0.24_{-0.13}^{+0.26}$     &  $0.55_{-0.29}^{+0.32}$   \\ 
$M_{2}$ $(M_{\rm J})$             &    $0.32_{-0.16}^{+0.34}$     &  $1.37_{-0.72}^{+0.80}$   \\ 
$a_{\bot}$ (au)                   &    $2.72_{-0.55}^{+0.49}$     &  $3.57_{-1.02}^{+0.68}$   \\ 
$D_{\rm L}$ (kpc)                 &    $7.07_{-1.15}^{+0.93}$     &  $6.47_{-1.73}^{+1.04}$    
\enddata 
\vspace{0.05cm}
\label{table:phys}
\end{deluxetable}

\end{document}